\documentstyle[12pt]{article}
\def\Z{{\bf Z}}

\def\J{{\cal J}}

\def\I{{\cal I}}
\def\F{{\cal F}}
\def\M{{\cal M}}

\date{}
\begin{document}
\title{{\LARGE Zero-Temperature Dynamics of Ising Spin Systems Following
a Deep Quench:  Results and Open Problems}} 
\author{
{\bf C. M. Newman}\thanks{Partially supported by the 
National Science Foundation under grant DMS-98-02310.}\\
{\small \tt newman\,@\,cims.nyu.edu}\\
{\small \sl Courant Institute of Mathematical Sciences}\\
{\small \sl New York University}\\
{\small \sl New York, NY 10012, USA}
\and
{\bf D. L. Stein}\thanks{Partially supported by the 
National Science Foundation under grant DMS-98-02153.}\\
{\small \tt dls\,@\,physics.arizona.edu}\\
{\small \sl Depts.\ of Physics and Mathematics}\\
{\small \sl University of Arizona}\\
{\small \sl Tucson, AZ 85721, USA}
}
\maketitle
\begin{abstract}
We consider zero-temperature, stochastic Ising
models $\sigma^t$ with nearest-neighbor interactions and an initial
spin configuration $\sigma^0$ chosen from a symmetric Bernoulli
distribution (corresponding physically to a deep quench).  Whether
$\sigma^\infty$ exists, i.e., whether each spin flips only finitely
many times as $t \to \infty$ (for almost every $\sigma^0$ and
realization of the dynamics), or if not, whether every spin
--- or only a fraction strictly less than one --- flips infinitely
often, depends on the nature of the couplings, the dimension, and the
lattice type.  We review results, examine
open questions, and discuss related topics.
\end{abstract}
\small
\normalsize

\section{Introduction}
\label{sec:intro}

The behavior of different kinds of magnetic systems following a deep
quench comprises a central topic in the study of their nonequilibrium
dynamics.  Rigorous and nonrigorous results have been obtained on
different questions arising naturally in this context: the formation
of domains, their subsequent evolution, spatial and temporal scaling
properties, and related questions (for a review, see
Ref.~\cite{Bray}); the persistence properties at zero and positive
temperature \cite{St,De,DHP,MH,MS,NS99a}; the observed aging phenomena
in both disordered and ordered systems (see, e.g.,
\cite{aging1,aging2,aging3,aging4}); and many others.

In this paper we will summarize results on a very basic question,
whose answer is not only relevant to the questions mentioned above but
often naturally precedes them.  Put informally, consider a quench
from infinite temperature to zero temperature, and let the system then
evolve using standard Glauber dynamics.  Will the spin configuration
eventually settle down to a final state, or will it continue to evolve
forever (and if so, in what sense)?  Here we will concern ourselves
mostly with this question of approach to a final state in different
Ising spin models, and will not address, except briefly in the last
section, the properties of such final
states.  We now state the problem more precisely.

Consider the stochastic process $\sigma^t=\sigma^t(\omega)$ corresponding
to the zero-temperature limit of Glauber dynamics for an Ising model with
Hamiltonian,
\begin{equation} 
\label{eq:EA}
{\cal H}= -\sum_{\scriptstyle\{x,y\}\atop\scriptstyle\|x-y\|=1} J_{x,y} \sigma_x \sigma_y\ .
\end{equation}
Here $||\cdot||$ denotes Euclidean length.  For now $\sigma^t$ takes values
in ${\cal S}=\{-1,+1\}^{{\bf Z}^d}$, the space of (infinite-volume) spin
configurations on the $d$-dimensional hypercubic lattice, but later (see
Subsec.~\ref{subsec:odd}) we will examine other types of lattices as well.
The initial spin configuration $\sigma^0$ is chosen from a symmetric
Bernoulli product measure (denoted $P_{\sigma^0}$), corresponding
physically to a deep quench.  If the spin model is disordered, then the
transition rates depend on a realization ${\cal J}$ of the (i.i.d., unless
otherwise specified) random couplings $J_{x,y}$, with (common, unless
otherwise specified) distribution $\mu$ on ${\bf R}$.  The (continuous
time) dynamics is given by independent (rate 1) Poisson processes at each
$x$ corresponding to those times $t$ [think of these as clock rings at $x$]
when a spin flip ($\sigma_x^{t+0}=-\sigma_x^{t-0}$) is {\it considered\/}.
If the resulting change in energy is negative (or zero or positive), then
the flip is done with probability 1 (or 1/2 [determined, say, by a fair
coin toss], or 0).  We denote by $P_\omega$ the probability distribution on
the realizations $\omega$ of the dynamics.  For most of this paper we
consider only this single-spin-flip dynamics, but in the last section will
briefly discuss multi-spin-flip dynamics.  The joint distribution of ${\cal
J}$, $\sigma^0$, and $\omega$ will be denoted $P$.

A natural question in both the disordered and non-disordered models is
whether $\sigma^t$ has a limit (with $P$-probability one) as $t \to \infty$
or equivalently whether for every $x$, $\sigma_x^t$ flips only finitely
many times. More generally, one may call such an $x$ an $\F$-site ($\F$ for
finite) and otherwise an $\I$-site ($\I$ for infinite).  By
translation-ergodicity, the collection of $\F$-sites (resp., $\I$-sites)
has (with P-probability one) a well-defined non-random spatial density
$\rho_\F$ (resp., $\rho_\I$).  The densities $\rho_\F$ and $\rho_\I$ depend
only on $d$, $\mu$, and possibly lattice type, and of course satisfy
$\rho_\F + \rho_\I = 1$. We characterize the
triplet ($d,\mu$, lattice type) as being type $\F$ or $\I$ or $\M$ (for
mixed) according to whether $\rho_\F = 1$ or $\rho_\I = 1$ or
$0<\rho_\F,\rho_\I<1$.

In the remainder of the paper, we 
review results for different
$d$, different lattices, and a number of important special cases of
$\mu$.  In Sec.~\ref{sec:1d}, we present results on
one-dimensional chains for both homogeneous and disordered systems.
In Sec.~\ref{sec:ferrosquare} we consider the homogeneous ferromagnet
(or antiferromagnet) on the square lattice $\Z^2$ and show it is type
$\I$.  In Sec.~\ref{sec:continuous} we review models with continuous
disorder, and discuss a theorem whose consequence is that most such
systems of interest, including ordinary random ferromagnets and the
Edwards-Anderson (EA) spin glass \cite{EA}, are type $\F$ for
any $d$ and lattice type.  We also
show why this theorem implies that homogeneous ferro- and
antiferromagnets, on lattices (in any $d$)
where each site has an {\it odd\/}
number of neighbors, are now type $\F$.  In
Sec.~\ref{sec:pmj}, we discuss $\pm J$ spin glasses on
$\Z^2$ and other models with noncontinuous disorder on $\Z^d$
that are 
type $\M$.  In Sec.~\ref{sec:summary} we summarize our findings and
list a number of open problems.  We also discuss there
several situations not discussed in the bulk of the paper, including
positive temperature, multi-spin-flip dynamics and persistence.

\section{One-dimensional models}
\label{sec:1d}

In one dimension the analysis is particularly simple, and it is not
hard to show that: a) when the couplings all have the same magnitude
(regardless of sign, since in one dimension all such models can be
gauge-transformed to the uniform ferromagnet), the model is type $\I$;
b) disordered models with $\mu$ continuous are 
type $\F$; and c)
models in which the couplings can (with positive probability) take on
two or more discrete magnitudes are type $\M$. A
proof of the first two claims may be found in \cite{NNS}, and 
a proof of c) in \cite{GNS}.
Here we summarize the proof of b); the proofs of
a) and c) are simplifications of the 
corresponding $d=2$ proofs given in Secs.~\ref{sec:ferrosquare} and
~\ref{sec:pmj}, respectively,
and so
will not be given separately.
Modified arguments can be used to examine
the dynamical behavior for $\mu$'s with both a continuous
part and a {\it single\/} discrete magnitude; see
\cite{GNS} for details.

A sketch of the proof of b) is as follows. Consider a site $x$ such that
$|J_{x,x+1}|$ {\it strictly\/} exceeds the two neighboring coupling
magnitudes; because $\mu$ is continuous, such sites occur with positive
density.  It is clear that once $\sigma_x \sigma_{x+1} = \hbox{sgn}
(J_{x,x+1})$ (either initially in $\sigma^0$, or through a subsequent flip
of one of the two spins as determined by $\omega$), neither spin will flip
again, demonstrating already that $\rho_\F>0$.  By translation-ergodicity
of $P$, there will be (with $P$-probability one) a doubly infinite sequence
of such sites $x_n$ (with positive density).

Consider now the interval $\{x_{n-1}+1,x_{n-1}+2,\ldots,x_n\}$.  By
the preceding argument, there will be some time after which, both
$\sigma_{x_{n-1}+1}$ and $\sigma_{x_n}$ cease to flip.  After
that time we have a Markov process (restricted to the interval)
with a finite state space.  Because
$\mu$ is continuous, each flip within the interval will strictly lower
the energy, which (for fixed $\J$) is bounded below,
by some minimal amount.  The process must
therefore eventually reach an absorbing state in which all spins have
stopped flipping.  Because this argument applies to every such
interval, a continuous $\mu$ in one dimension is type $\F$.

\section{Homogeneous ferromagnet on $\Z^2$}
\label{sec:ferrosquare}

In \cite{NNS} it was shown that the homogeneous ferromagnet on $\Z^2$
is type $\I$; we will sketch the argument here.  Essentially the
same argument holds for the homogeneous antiferromagnet on $\Z^2$.  In
this section $P$ refers to the joint 
distribution of $\sigma^0$ and $\omega$.

To begin, we note that two possible
absorbing states are the two uniform spin configurations:
$\sigma_x \equiv +1$ for every $x$ and $\sigma_x \equiv -1$ 
for every $x$.  Because
of global spin-flip symmetry, these outcomes must have equal
probability $p$ $(\le 1/2)$.  But because of translation-invariance and
translation-ergodicity of $P$, $p$ must be either 0
or 1; therefore, $p=0$.
The only other absorbing states are those with one or more parallel
domain walls separating strips of uniform $+1$ and $-1$ spin
configurations; if the state is to be absorbing, then all these
domain walls must be parallel to either the $x$- or $y$-axis.  Using
the argument above, but with spin-flip symmetry
replaced by invariance with respect to rotations by
$\pi/2$, we conclude similarly that each of these 
two sets of outcomes must also
have zero probability.  This shows at least that $\rho_\I>0$.

To show that the model is type $\I$, we need to show that
{\it every\/} spin flips infinitely often.  By 
translation-invariance and spin-flip symmetry, if
$\rho_\F>0$, then the following must occur with positive
$P$-probability: for some $x=(x_1,x_2)$,
$y=(x'_1,x_2)$ and $z=(x_1,x'_2)$ with $x_1 < x'_1$ and $x_2 < x'_2$
and for some $t'$, $\sigma_x^t = +1$ and $\sigma_y^t = -1$ and
$\sigma_z^t = -1$ for all $t \ge t'$. But this would require that
\begin{equation}
\label{eq:cond2}
\inf _{\sigma \in {\cal S}''}P_{\omega}
(\sigma^{t+1}\notin{\cal S}''| \sigma^t = \sigma) = 0,
\end{equation}
where ${\cal S}''$ is the set of spin configurations on ${\bf Z}^2$
with the values $+1, -1, -1$ at the sites $x, y, z$. 

But this is not so as can be seen from the following argument.
Let us define $w=(x'_1,x'_2)$ and ${\cal R}$ to be the rectangle
with corners at $x,y,z$ and $w$. 
Any spin configuration in ${\cal S}''$ must have a
domain wall 
{\it that is contained within\/} ${\cal R}$
and that either (a) connects the $[x,y]$ straight
line segment to the $[x,z]$ segment or else (b) connects the 
$[x,y]$ segment to the $[y,w]$ segment or else (c) connects the 
$[x,z]$ segment to the $[z,w]$ segment. 
In case (a), there is some sequence of clock rings
(and absence of rings) and coin toss outcomes within  
${\cal R}$ during the time interval
$[t,t+1]$ (that occurs with $P_{\omega}$-probability
bounded away from zero) that will move the domain wall
towards the Southwest so that $\sigma_x^{t+1} = -1$.
Similarly, in cases (b) or (c), the domain wall can 
move to the Southeast or to the Northwest so that 
$\sigma_y^{t+1} = +1$ or $\sigma_z^{t+1} = +1$. 

\section{Models with continuous disorder}
\label{sec:continuous}

A central result is that in any dimension (and on any lattice),
models with continuous disorder 
distribution $\mu$ of finite mean are type $\F$. 
These models include EA spin glasses with a Gaussian $\mu$ and
random ferromagnets with a uniform $\mu$.
The idea behind the proof was already used in Sec.~\ref{sec:1d}
as part of the proof that one-dimensional models with continuous
$\mu$ are type $\F$.  Here we sketch the more general proof;
for further details, see \cite{NNS}.  

 Let $\sigma_x^t$ be the value of $\sigma_x$ at time $t$ 
for fixed $\omega$, $\sigma^0$ and ${\cal J}$.  Let 
\begin{equation}
\label{eq:energy}
E(t)=-(1/2)\overline{\sum_{y:||x-y||=1}J_{xy}\sigma_x^t\sigma_y^t}\, ,
\end{equation}
where the bar indicates an average over $P$.
By translation-ergodicity of $P$,
and using the assumption that
$\overline{|J_{xy}|} < \infty$,
it follows that $E(t)$ exists, 
is independent of $x$, and equals the energy density (i.e., the
average energy per site) at time $t$ in almost every realization of ${\cal
J}$, $\sigma^0$, and $\omega$.

Because every spin flip lowers the energy, $E(t)$ monotonically decreases
in time (note that $E(0)=0$) and has a finite limit $E(\infty)$ ($\ge
-d\overline{|J_{xy}|}$).  Now choose any fixed number $\epsilon>0$, and let
$N_x^\epsilon$ be the number of spin flips (over all time) of the spin at
$x$ that lower the energy by an amount $\epsilon$ or greater.  Then
$-\infty<E(\infty)\le -\epsilon\overline{N_x^\epsilon}$ so that for every
$x$ and $\epsilon > 0$, $N_x^\epsilon$ is finite.  Let $\epsilon_x$ be the
minimum energy (magnitude) change resulting from a flip of $\sigma_x$; then
although $\epsilon_x$ varies (differently in each ${\cal J}$) with $x$, it
is sufficient that it is strictly positive.

It is implicit in the proof of this result that, even without the
continuity assumption on the distribution of couplings, with
$P$-probability one there can be only finitely many flips of any spin
that cause a {\it nonzero\/} energy change.  When $\mu$ is continuous,
the probability of a ``tie'' in any sum or difference of a given spin's
nearest-neighbor coupling strengths (and therefore the probability of
a spin flip costing zero energy) is zero, yielding the result that
these systems are type $\F$.

\subsection{Homogeneous ferromagnets on lattices other than $\Z^d$}
\label{subsec:odd}

It follows from the argument above that this type $\F$
result applies also to
homogeneous Ising spin systems on lattices with an {\it odd\/} number
of nearest neighbors, so that ties in energy cannot occur.  
(It also applies to $\pm J$ models (as defined below) on
these lattices.) Such
lattices include the two-dimensional hexagonal (or honeycomb) lattice,
and the double-layered cubic lattices $\Z^d\times \{0,1\}$ (i.e., a
``ladder'' when $d=1$, two horizontal planes separated by unit
vertical distance when $d=2$, and so on).

\subsection{Continuous disorder with infinite mean}
\label{subsec:infinite}

What about models where the disorder is continuous but the mean is
infinite?  We can show that a restricted class of these models are 
type $\F$; these are models where {\it influence percolation\/}
\cite{NN} does not occur.  Such models include all one-dimensional
models with continuous disorder (regardless of whether the mean is
finite), and strongly and highly disordered models \cite{NS94,BCM,NS96}.
We refer the reader to \cite{NNS} for a discussion of this situation.

\section{$\pm J$ and related models}
\label{sec:pmj}

We now turn to models on $\Z^d$ where
\begin{equation}
\label{eq:discretemu}
\mu=\alpha\delta_{J_1}+(1-\alpha)\delta_{J_2}\ , 
\end{equation}
with $0<\alpha<1$ and $J_1\ne J_2$.  When $J_1=-J_2 \neq 0$, 
we call this a $\pm J$ model.
(In much of the literature, $\pm J$ {\it spin
glasses\/} refer to the specific case $\alpha=1/2$.)
These models have been analyzed in
\cite{GNS}, along with modifications where, e.g., $\mu$
consists of both a continuous and a discrete part.  Because such
modified models appear to be of less interest, we note here
only that these models are generally
type $\M$, and refer the reader to
\cite{GNS} for details; for the remainder of this section, we confine
ourselves to distributions of the form given in
Eq.~(\ref{eq:discretemu}).

The main results of \cite{GNS} were to prove the following two
assertions:  a) models in which $J_1\ne -J_2$ are type $\M$ in
any dimension, and b) $\pm J$ models in two dimensions are also
type $\M$.  The first of these is much easier to prove, and we start
with that case.

We sketch here a proof for the $d=2$ case; the extension to other $d$ is
straightforward.  To show that $\rho_\F>0$ on $\Z^2$, let $|J_1|<|J_2|$ and
consider a plaquette all of whose edges have coupling $J_2$, and all edges
outside the plaquette but that connect to one of its corners have coupling
$J_1$.  With positive $P$-probability, such a coupling configuration will
occur with the spins at the four corners initially (or eventually) all $+1$
(or all $-1$) if $J_2 >0$ or else alternating in sign if $J_2 < 0$.  These
spins will thereafter never flip, proving that $\rho_\F>0$.

To show that $\rho_\I>0$, consider a configuration in which the couplings
between some $w$ and its four neighbors $z_i$ are all $J_1$. Suppose also
that each $z_i$ belongs to a plaquette satisfying the set of conditions
described in the previous paragraph, with the spins at $z_1$ and $z_2$
equal to $+1$ and the spins at $z_3$ and $z_4$ equal to $-1$.  This will
ensure that the spin at $w$ flips infinitely often, and because such a
situation will occur with positive $P$-probability, $\rho_\I>0$.

We now turn to a discussion of b).  It is relatively easy to show that
for two-dimensional $\pm J$ models, $\rho_\I>0$, and we will sketch the
proof of that here.  It takes considerably more work to show that
$\rho_\F>0$, and so we will present here only the idea behind the proof,
and refer the reader to \cite{GNS} for details.

To show that $\rho_\I>0$, we consider a configuration of a $5 \times 5$
square of sites in the dual lattice $\Z^{2*} \equiv \Z^2 +(1/2,1/2)$ in
which exactly $9$ of the $25$ sites are frustrated (corresponding to
plaquettes in $\Z^2$ with an odd number of negative couplings): the central
site $w_c$; two Southeastern sites, $w_1 = w_c + (1,-1)$ and $w_2 = w_c +
(2,-1)$; and six other sites obtained from $\{w_1, w_2\}$ by (multiple)
$\pi/2$-rotations about $w_c$. Such a frustration configuration occurs with
positive $P$-probability, so we are done if we can show that there always
exists at least one $\Z^2$ site (among the $36$ whose plaquettes form our
$5 \times 5$ square) that has positive flip rate.  We now proceed to do
this.

We note first that any domain wall (i.e., a path of unsatisfied edges in
$\Z^{2*}$ [connecting a pair of frustrated sites]) that is not straight
implies a site in $\Z^2$ with positive flip rate; this is because there
must exist a $\Z^2$ site with at least two unsatisfied edges.  Now in our
frustration configuration, there must be a domain wall starting from $w_c$.
Either this domain wall already determines a positive flip rate site
because it is not straight, or else it runs straight out of the square; in
the latter case, by the invariance with respect to rotations by $\pi /2$,
we may assume (without loss of generality) that the domain wall emanating
from $w_c$ runs to the East and passes just above the (dual) edge joining
the two Southeastern sites, $w_1$ and $w_2$. But then there must be another
domain wall starting from $w_1$.  Either these two domain walls together
determine a positive flip rate site or else the second one runs from $w_1$
straight out of the square to the South. But then there must be a third
domain wall starting from $w_2$, that (together with the previous two) will
determine a positive flip rate site, no matter what direction it runs off
to.  Using as usual translation-invariance and translation-ergodicity of
$P$, we conclude that $\rho_\I>0$.

A proof demonstrating that $\rho_\F>0$ is considerably more involved,
as noted, but the general strategy is similar.  We again
consider an event involving the frustration configuration in a finite
region of $\Z^{2*}$, and the spin configuration in a related region of
$\Z^2$.  One wants to show that at least one of these $\Z^2$ sites 
will {\it eventually\/} have flip rate zero and hence will flip
only finitely many times, thus proving $\rho_\F>0$.  This is done by
proving that the domain wall geometry in $\Z^{2*}$ must eventually
satisfy various constraints, in particular that certain contour events
recur indefinitely with probability zero; otherwise, there would
be infinitely many energy-lowering flips in the fixed square with
positive probability, violating the result presented in
Sec.~\ref{sec:continuous}.  For further details, see \cite{GNS}.

\section{Summary and open problems}
\label{sec:summary}

We have studied the dynamical evolution of several categories of Ising
spin models, both ordered and disordered, following a deep quench
(from infinite to zero temperature), in all dimensions and for
different kinds of (regular) lattices.  Our concern here has centered
on the question of existence of a final state, given the usual
zero-temperature Glauber dynamics and the Hamiltonian~(\ref{eq:EA}).
It is interesting to consider other types of situations, but these
will be left for the future.  These include, for example, other kinds
of spin models (Potts, XY, Heisenberg), initial spin
configurations not chosen from the symmetric Bernoulli distribution,
and others.  Nevertheless, we feel that substantial progress has so
far been made, and we review the results below, including a discussion
of remaining open questions (for this type of model and situation).

\subsection{Review of results}
\label{subsec:review}

All results below are for Ising spin systems with
Hamiltonian~(\ref{eq:EA}), and ${\cal J}$, $\sigma^0$ and $\omega$ chosen as
discussed in Sec.~\ref{sec:intro}.  A given system may have three
possible dynamical outcomes: it may be type $\I$, $\F$, or
$\M$, whose meanings were given in Sec.~\ref{sec:intro}.

\medskip

\noindent {\it Homogeneous ferromagnets and antiferromagnets\/}: In
one dimension and in two dimensions on $\Z^2$, these are type $\I$.
In any dimension on a lattice where each site has an odd number of
nearest neighbors, they are type $\F$ \cite{NNS}.

\medskip

\noindent {\it Models with continuous disorder\/}: These need to be
further subdivided.  The most important, and commonly studied, cases
are those models where the coupling distribution has finite mean;
these include ordinary (EA) spin glasses and random ferromagnets.  In
all dimensions (and for all lattices)
these models are type $\F$ \cite{NNS}.  Another class
of models that are type $\F$ \cite{NNS} are those in which influence
percolation \cite{NN} does not occur; these include all
one-dimensional models with continuous disorder, and the strongly and
highly disordered models of spin glasses and random ferromagnets
\cite{NS94,BCM,NS96}.

\medskip

\noindent $\pm J$ {\it models\/}: In one dimension, these are
type $\I$; in two dimensions, type $\M$ on the square
lattice \cite{GNS}.  On a lattice where each site has an odd number of
neighbors, they are type $\F$ in any dimension.  

\medskip 

\noindent {\it Models with other\/} $\mu$:  
If $\mu$ is of the form
$\alpha\delta_J+\beta\delta_{-J} +\nu$ with $J>0$, $0<\alpha+\beta<1$,
and $\nu$ continuous and supported on $[-J,+J]$, then it is type $\F$
in one dimension \cite{GNS}.
Other examples are distributions
of the form Eq.~(\ref{eq:discretemu}), where $|J_1|\ne |J_2|$,
distributions supported partially on a continuous interval and
partially on atoms at discrete values, and so on.  These are type $\M$
in all dimensions \cite{GNS}.

\subsection{Open problems}
\label{subsec:open}

\noindent

For Ising spin systems with Hamiltonian~(\ref{eq:EA}), and 
${\cal J}$, $\sigma^0$
and $\omega$ chosen as discussed in Sec.~\ref{sec:intro}, the problems
of the dynamical outcomes of the following situations remain open:

\medskip

\noindent 1) Homogeneous ferromagnets (and antiferromagnets) on the
lattice $\Z^d$ (or others except where each site has an odd number of
neighbors) for $d\ge 3$.  Numerical results \cite{St} suggest that
these should be type $\I$ for $d=3$ and perhaps $4$, but
may possibly be type $\F$ (or
$\M$) for $d\ge 5$.  

\medskip

\noindent 2) Models with continuous disorder in $d>1$, where both the
coupling distribution has infinite mean and influence percolation
occurs.  The proof \cite{NNS} sketched in Sec.~\ref{sec:continuous}
already implies that $\rho_\F>0$ for these models, so they are either
type $\F$ or $\M$.  It seems 
reasonable to conjecture that they are type $\F$.

\medskip

\noindent 3) $\pm J$ models on $\Z^d$ (or other lattices except where
each site has an odd number of neighbors) for $d\ge 3$.  There is
little we can say about these right now, except that 
it would be interesting to relate $\pm J$ models to homogeneous
ferromagnets on the same lattice. E.g., perhaps
$\rho_\F(\pm J$ model)$\ge
\rho_\F$(homogeneous ferromagnet).  
If this were the case, then
demonstrating that homogeneous ferromagnets are type $\F$ on $\Z^d$
for $d\ge 5$ would immediately resolve the question for $\pm J$ spin
glasses on those lattices.

\subsection{Related topics}
\label{subsec:related}

In this section we touch on several related topics.  In particular,
throughout this paper we have concerned ourselves solely with the
question of convergence of $\sigma^t$ (with $P$-probability one) to
a final state, but have not examined the properties of this final
state (when it exists) in the different models of interest.  We also
have not discussed rates of convergence to the final state.  We will
briefly discuss these issues, but first will consider 
positive temperature.

\medskip

\noindent {\it Positive temperatures\/}.  Here one 
treats the
behavior of the local order parameter rather than that of single
spins.  Construction of dynamical measures, analysis of their
evolution, and relation to pure state structure are extensively
discussed in Ref.~\cite{NS99b}.  Here we mention only
a few relevant results.

It should first be noted that the categorization into types $\I$,
$\F$, and $\M$ needs to be modified and refined at positive
temperature.  Without going into detail, we will simplify matters here
by dividing systems into those where on any finite lengthscale, the
system equilibrates (into a pure state) after a finite time (depending
on $\sigma^0$, $\omega$, the lengthscale, and ${\cal J}$ if relevant),
in the sense that interfaces cease to move across the region after
that time; and those where this local equilibration does not occur.
The latter case we call {\it local non-equilibration\/} (LNE), of
which there are two types.
A precise definition requires the use of a dynamical measure; we
refer the reader to \cite{NS99b} for details and to \cite{FIN}
where one of the types ({\it Chaotic Time Dependence\/}) is
shown to occur in a $d=1$ model with disordered rates.

A main result of \cite{NS99b} is that if only a single pair, or
countably many pairs (including a countable infinity) of pure states
exists (with fixed ${\cal J}$), and these all have nonzero EA order
parameter \cite{EA}, then LNE occurs.  A corollary is that if LNE does
{\it not\/} occur, and the limiting pure states have nonzero EA order
parameter, then there must exist an {\it uncountable\/} infinity of
pure states, with almost every pair (as the realizations of the
initial state and dynamics vary) having overlap zero.

One consequence of these results is that LNE (in a rough sense the
positive temperature equivalent of type $\I$ behavior) 
occurs at
positive temperature (with $T<T_c$) in the $2D$ uniform ferromagnet
and (presumably) random Ising ferromagnets for $d<4$.  Because the
number and structure of pure states at positive temperature in Ising
spin glasses is unknown for $d\ge 3$ (and, from a rigorous point of
view, unproved even for $d=2$), occurrence of LNE there remains
an open question.

\medskip

\noindent {\it Multi-spin dynamics\/}.  In Ref.~\cite{NS99c} we examined
ordinary spin glasses and random ferromagnets in arbitrary dimension and
considered an extension of the zero-temperature single-spin flip Glauber
dynamics in which rigid flips of all lattice animals (i.e., finite
connected subsets of $\Z^d$, not necessarily containing the origin) up to
size $M$ spins can occur.  Of particular interest here is the limit
$M\to\infty$.  For the dynamics to remain sensible, we need to choose the
rates for $K$-spin lattice animal flips to decrease as $K$ increases, so
that the probability that any fixed spin considers a flip in a unit time
interval remains of order one, uniformly in $M$. A further requirement for
the dynamics to be well-defined is that information not propagate
arbitrarily fast throughout the lattice as $M\to\infty$.  Such a dynamics
can be constructed, and generates infinite-volume ground states
\cite{NS99c}.

\medskip

\noindent {\it Persistence\/}.  A topic of current interest is the
$P$-probability $p(t)$ at time $t$ that a spin has not yet flipped.  For
the homogeneous ferromagnetic Ising model on $\Z^d$, this probability has
been found to decay at large times as a power law $p(t)\sim t^{-\theta(d)}$
\cite{St,De,DHP} for $d<4$.  The ``persistence'' exponent $\theta(d)$ is
considered to be a new universal exponent governing nonequilibrium dynamics
following a deep quench \cite{MS}.  Our work \cite{NS99a} shows that
dependence on lattice type (e.g., square vs.~hexagonal lattices for
homogeneous ferromagnets) implies nonuniversality of this behavior, and
moreover that the persistence phenomenon (which seems to require a system
to be type $\I$) is unstable to the introduction of randomness into the
spin couplings.  Moreover, in some simple systems it was shown that the
decay $p(t)-p(\infty)$ to the final state is {\it exponential\/} as
$t\to\infty$.

\medskip

\noindent {\it Properties of the final state.\/} We turn finally to a
discussion of the properties of the limiting state $\sigma^\infty$ for
ordinary spin glasses and random ferromagnets in any dimension.  For
the $M$-spin-flip dynamics (with $M$ finite) discussed above, the
final states are energetically stable up to a flip of any subset of
$M$ spins; we call these $M$-spin-flip stable states.  A number of
results were obtained in \cite{NS99c}; a central result is that
there is an uncountable infinity of these metastable states, in any
$d$ and for any $M$, and their overlap distribution is a delta-function
at zero.  For further results and discussion, we refer the reader to
\cite{NS99c}.

\newpage

\bigskip
\bigskip
\noindent

\medskip

\renewcommand{\baselinestretch}{1.0}

\end{document}